# Gate-tunable emission of exciton-plasmon polaritons in hybrid MoS$_2$-gap-mode metasurfaces

*Peinan Ni[1, ‡], Andrès De Luna Bugallo[2, ‡, *], Victor M. Arellano Arreola[2], Mario Flores Salazar[2], Elodie Strupiechonski[2], Virginie Brändli[1], Rajath Sawant[1], Blandine Alloing[1], and Patrice Genevet[1, *]*

[1]Université Cote d'Azur, CNRS, CRHEA, rue Bernard Gregory, Sophia Antipolis 06560 Valbonne, France

[2]CONACYT—Cinvestav Unidad Querétaro, Querétaro, Qro. 76230, Mexico

**ABSTRACT:** The advance in designing arrays of ultrathin two-dimensional optical nano-resonators, known as metasurfaces, is currently enabling a large variety of novel flat optical components. The remarkable control over the electromagnetic fields offered by this technology can be further extended to the active regime in order to manipulate the light characteristics in real-time. In this contribution, we couple the excitonic resonance of atomic thin MoS$_2$ monolayers with gap-surface-plasmon (GSP) metasurfaces, and demonstrate selective enhancement of the exciton-plasmon polariton emissions. We further demonstrate tunable emissions by controlling the charge density at interface through electrically gating in MOS structure. Straddling two very active fields of research, this demonstration of electrically tunable light-emitting metasurfaces enables real-time manipulation of light-matter interactions at the extreme subwavelength dimensions.

**Introduction**

Two-dimensional (2D) materials featuring exceptional optical and optoelectronic properties have attracted increasing interest with significant potential to develop a great number of emerging nano-optoelectronics devices.[1-4] In particular, 2D materials host a larger variety of polaritons compared with conventional three-dimensional (3D) semiconductors, such as plasmon polaritons in graphene,[5,6] phonon polaritons in hexagonal boron nitride (hBN).[7,8] Moreover, the current "polaritonic library" of 2D materials has been greatly expanded by the wide observation of exciton polaritons from the recent emerging atomically thin semiconductors, such as molybdenum and tungsten-based transition metal dichalcogenides (TMDCs).[9-11] Exciton polaritons are new quasi-particles of light-matter hybrid states, which both have the advantages of photons, such as ultra-fast speed, long-range spatial/temporal coherence, and exhibit strong light matter interactions. Their unique half-light half-matter characteristics make them indispensable platform to both explore fundamental research such as polariton Bose-Einstein condensation,[12,13] polariton bistability,[14-16] and propose new types of polaritonic devices like polariton lasers,[12,17] switches,[18] transistors,[14] and logic gates.[15,16] Among TMDCs, monolayer molybdenum disulfide ($MoS_2$) has been one of the most widely studied. Various excitons and other types of exciton complexes such as trions and biexcitons can be readily observed and manipulated even at room temperature in MoS$_2$ monolayer due to its large exciton binding energies, which together with its direct electronic band structure, make it a unique system to investigate polaritonic phenomena.[19-22]

Over the past 50 years, the field of plasmonic has received considerable attention, in particular due to its extraordinary capability of confining light below the diffraction limit. Recently, exciton-plasmon coupling has been extensively observed from a large variety of hybrid 2D material plasmonic systems, including metallic nanoparticle arrays, nanoslits and nanolattice.[23-26] Such coupling offers an additional degree of freedom to tailor the optical properties of 2D materials, such as quantum efficiency enhancement, polarization control, and so forth. Moreover, the feasibility of controlling of the exciton-plasmon coupling through electrical gating has been recently demonstrated, opening up great opportunities to modulate light-matter interactions inside 2D materials in real-time.[27] On the other hand, the emerging two-dimensional ultra-thin optical components, known as metasurfaces, exhibit exceptional spectral and spatial manipulation of electromagnetic waves within subwavelength dimensions, which provide powerful tools to mold the light in a very compact and efficient way.[28-30] Further extending this remarkable controllability of light-matter interactions into the active regime might eventually lead to various functionalized ultra-compact active devices.

In this work, gap-surface-plasmon (GSP) metasurfaces are designed and coupled with selected exciton resonance of monolayer $MoS_2$ to enhance exciton-plasmon polariton emissions with controlled polarization. Furthermore, gate-tunable exciton plasmon polariton emissions are realized in $MoS_2$ material under metal-oxide-semiconductor (MOS) configuration for the first time. The tunability relies on the dynamic tuning\detuning process between the neutral exciton resonance and the trion resonance inside $MoS_2$ material enabled by the manipulation of charge density at interface under applied gate bias. Moreover, this work demonstrates the capability of manipulating light-matter interactions in the active regime at the subwavelength dimensions thought metasurface technologies, opening up important opportunities to develop a wide range of lightweight, ultra-compact and power efficient optical components and optoelectronic devices.

**Results and discussion**

Both atomic force microscopy (AFM) and Raman spectroscopy are used to identify the monolayer thickness of the $MoS_2$ material obtained in this work, as shown in Fig. 1(a). The growth details can be found in the supplementary information. The trigonal prismatic coordination of a monolayer of $MoS_2$ as shown in inset Fig. 1a is known to be a direct band semiconductor with the highest valence band and the lowest conduction band formed primarily from the Mo d-orbital. We begin the design of the hybrid $MoS_2$ metasurface by analyzing the light emission properties of the as-grown $MoS_2$ monolayers on the $SiO_2$/Si substrate, based on their photoluminescence (PL) spectra, as shown in Fig. S1. By decomposition the PL emissions through a Lorentzian fitting, three PL components around 620 nm, 676 nm and 686 nm are revealed, respectively. It is well-demonstrated that by approaching an emitter close to an infinitely planar gold (Au) film, the latter known for supporting strong surface plasmon resonance (SPR) in the visible range, the spontaneous photoluminescence decay rate of the emitter, denoted by $\Gamma_{rad}$, will be significantly fasten due to the access to new radiative pathways. Interestingly, such system, which is supposed to feature high spontaneous emission rate, often fails in further enhancing the radiative efficiency whenever emitters are brought in very close proximity to metallic films. The physical mechanisms responsible for quenching of the photoluminescence properties are first related to the unavoidable dissipation effects arising from the strong electromagnetic field confinement and local field enhancements in lossy metal but also from the inelastic scattering of electrons close to the Fermi surface.[31] In order to overcome these limitations, it is critical to reduce the numbers of non-radiative decay channels. Since the inelastic terms strongly depends on the proximity of the emitter with respect to the metallic surface, we could eventually benefit from the strong field enhancement, *i.e.* improving $\Gamma_{rad}$, meanwhile decreasing quenching effects from inelastic coupling with electrons in the Fermi sea by controlling the distance of the MoS$_2$ emitters with respect to the gold interface. To this end, a 100 nm SiO$_2$ was sputtered onto the Au films before the transfer of MoS$_2$. Note that the thickness of the SiO$_2$ dielectric layer is chosen based on the tradeoff between the optimization of the electromagnetic field enhancement and reducing the leakage current for effective gate bias modulations. The use of other high index materials with reduced thickness, such as HfO$_2$, is expected to further improve the performance of the structure due to the larger plasmonic near-field enhancement effect. By doing this, it is found that excitonic features of the MoS$_2$ have been considerably enhanced with blue-shifts, as shown in Fig. 1b. To identify the different emission channels of the MoS$_2$ transferred onto the SiO$_2$/Au substrate, we performed a Lorentzian fitting of the photoluminescence and revealed three resonances with enhanced intensity around 613 nm, 650 nm and 674 nm. The first two features are known as the A and B excitons. They are associated with direct optical transitions from the highest spin-split valence band to the lowest conduction band at the K point of the Brillouin zone. The formation of these bound excitons transitions can be significantly modified by strong Coulomb interactions, leading to bound states of two electrons and a hole (trion states), creating an additional resonance around 674 nm. The origins of the excitonic transitions in monolayer MoS$_2$ transferred onto SiO$_2$/Au substrates were further confirmed by excitation-dependent measurements with pumping powers ranging from 10 to 500 μW at room temperature as shown in Fig. 1c. By employing the power law expression: *I α P$^γ$*, where *I* is integrated PL intensity and *P* is the laser power excitation, the nature of the excitonic processes can be identified. Fitting by a nonlinear regression model, as shown in Fig. 1d, the gamma coefficients of both the peak located at 650 nm and the peak around 613 nm are determined to be 1, which suggest exciton-like transition (exciton A and exciton B) while the value of the feature peaked at 674 nm is found to be 1.3, confirming the presence of a trion. These initial luminescence characterizations show that the SPR in gold film can significantly tailor the emission spectrum of the MoS$_2$ in the proposed hybrid structure, including 1) enhance the emission peaks intensity, and 2) modify the shape, in particular the positions, of the emission peaks due to the coupling of the plamonic resonance and the excitonic resonances of MoS$_2$, which have been widely observed from the plasmon/MoS$_2$ integrated systems.[32-35] Moreover, there is another advantage of the introduction of the underlying Au mirror as it produces multiple reflections which will increase the overall light absorption of the MoS$_2$ flake as shown in previous works.[36] However, such absorption enhancement does not present strong wavelength dependence and will not change the components of emissions, which is different from the selective enhancement caused by exciton-plasmons coupling proposed herein.

It is worth noticing that the various excitonic emissions from the MoS$_2$ enhanced by Au film open up unique opportunities to explore and develop ultra-compact tunable exciton-plasmon polariton light source by manipulating the coupling between excitonic resonance and SPR inside this structure. To this end, a tunable polaritonic light emitting structure is designed and fabricated considering an ultra-compact hybrid MoS$_2$-gap-mode metasurface by integrating metallic nanostrip-array onto the MoS$_2$ flake, as shown in Fig. 2a. In this structure, each metallic Au nanostrip can be considered as individual gap plasmon cavities. Their subwavelength arrangement constructs a metasurface that introduces a very basic polarization manipulation property. The width of nanostrips need to be designed to establish a gap plasmons resonance effect given by the standard Fabry-Pérot resonance formula:[37]

$$wk_0 n_{gsp} + \phi = m\pi$$

Where w is the width of the nanostrip, $k_0$ is the vacuum wave number, $n_{gsp}$ is effective index of the GSP, m is an integer defining the mode order, and ϕ is an additional phase shift acquired by the gap plasmon upon reflection at the edges of the metallic top covering strip. As discussed by M. G. Nielsen *et al*,[37] the non-zero phase shift depends on structural and material parameters and increases with increasing gap layer towards $\phi \sim \frac{3}{5}\pi$ for relatively thick cladding layer of about 100nm. Taking these estimations into account, the effective mode index of our gap plasmon mode is estimated at $n_{gsp} \sim 1.8$. Finite difference time domain simulations showing the reflectance of the bare GSP metasurface are performed to identify the geometrical parameters to achieve the proper overlap between the resonance and the PL peak of MoS$_2$ emitters. We focus on the emissions of exciton A considering its larger spectral weight. To interact efficiently with this resonance, the nanostrip resonance of the metasurface is designed at around 650 nm, matching with the peak of exciton A emissions, as shown in Fig. 2b. At resonance, the exciton energy can be effectively coupled with the GSP mode, giving rise to exciton-plasmon polariton. The close proximity of the nanostrips and bottom metallic mirror creates a tight optical confinement which could enhance the light emission properties, as indicated by the full electromagnetic wave simulations presented in Fig. 2c. Field enhancement confines the electromagnetic energy mostly within the gap region in a mode volume in the order of $V_{eff} \sim 3.5 \; 10^{-4} \; \mu m^3$, almost two order of magnitude smaller than the diffraction limited mode volume $\left(\frac{\lambda}{2n_{gsp}}\right)^3$ considering λ as the free space PL wavelength.[38] Large confinement factor provided by the gap plasmon nanocavities will tremendously enhances the above mentioned radiative quantum efficiency of the active emitters that are placed in the large electromagnetic field enhancement regions due to the Purcell effect. To benefit from this effect, $MoS_2$ emitters are directly templated on the cladding layer and brought in contact to the top metallic stripes that have been thereafter fabricated and aligned using standard electron beam lithography alignment accordingly to the $MoS_2$ positions. Note that with our fabrication process, $MoS_2$ is brought in direct contact to the top metallic layer, thus enabling electrical gating but unfortunately inducing coupling of the emitters with the top metallic Fermi electrons responsible for additional loss due to inelastic electron scattering.

Figur 3a shows the scanning electron micrographs (SEM) of the fabricated hybrid $MoS_2$-gap-plasmon metasurfaces. As can be seen from Fig. 3b, the emissions around 650 nm have been significantly enhanced due to the tight electromagnetic confinement at the resonance of the hybrid gap plasmon metasurfaces, which is in good agreement with the on-resonance design of the exciton A-plasmon coupling. Furthermore, direct comparison of PL of the same $MoS_2$ flake placed on $Au/SiO_2$ at the interface with the cavity at is provided in Fig S2, confirming strong enhancement due to gap plasmon coupling. Although significant enhancement has been observed from this structure, part of the emissions must have been quenched by absorption at the Au stripes given that the $MoS_2$ flake is in direct contact with top electrodes. One can imagine that the emission enhancement would be further increased to a large extent by inserting a dielectric layer. For example, one could encapsulate the MoS$_2$ in between atomically smooth layers of 2D hexagonal boron nitride (h-BN) as done routinely to further improve the carrier distribution and the MoS$_2$ emission properties. Furthermore, the strong anisotropy introduced by the metasurface design provides a feasible and effective approach to control the polarization of light emission, as shown in Fig. 3c. It is found that emission intensity of the TM component (transverse magnetic, electric field component perpendicular to the Au stripes) is almost an order of magnitude stronger than the TE component (transverse electric, electric field component parallel to the Au stripes). The polarization anisotropy can be characterized by the polarization ratio $\rho = (I_{TM} - I_{TE})/(I_{TM} + I_{TE})$, leading to a polarization ratio of ~0.52. Figure 3d shows the corresponding polar plot of the emission intensity. The polarization dependent emission from this hybrid structure further confirms the plasmonic nature of the enhanced emission. Moreover, it also suggests that other types of metasurface designs such as zigzag wires, asymmetric metallic rods or chiral structures could be employed to further manipulate the polarization properties. Valley-polarized photoluminescence in MoS$_2$ have been already tailored for example using chiral metasurfaces, spin-dependent plasmonic surface or asymmetric gradient plasmonic surfaces.[39-41]

In addition, it has been demonstrated that the various excitonic features presented on diverse 2D material can be simply manipulated electrically, optically and chemically by controlling the density and polarity of the carriers. [42-44] The electrical doping based modulation mechanism in field effect transistor configuration (FET) offers a fast and power-efficient approach to create a new type of ultra-compact tunable optoelectronic devices based on TMD materials. As a proof of such concept, gate tunable exciton-plasmon polariton light sources are proposed and realized in hybrid $MoS_2$-gap-plasmon metasurfaces, in which the Au nanostrip together with its underneath $MoS_2$ monolayer, $SiO_2$ spacing layer and the bottom Au mirror form an effective metal-oxide-semiconductor (MOS) capacitor as shown in the inset of Fig. 4a). In this design, the nanostrip acts as a ground, the bottom Au mirror serves as a gate, and the $SiO_2$ layer exhibits good dielectric characteristic, as demonstrated in Fig. S3. Upon application of an electrical bias between the nanostrip ground and the underlying gate plane, the free electrons concentration at the $MoS_2/SiO_2$ interface will be changed to either an electron accumulation or a depletion layer. A $MoS_2$ based FET device with the same thickness of $SiO_2$ dielectric layer was fabricated to estimate the change of electron density as a function of gate voltages, as shown in Fig. S4. The formation of excitons in the $MoS_2$ is modified by changing the electron density under different gate voltages, which will modulate the light emissions accordingly, as observed in Fig. 4(a). This modulation mechanism can be well understood in term of the band diagram: Positive gate bias can drive excess electrons to the interface between $SiO_2$ and $MoS_2$ creating an electron accumulation layer at the interface (Fig. 4c). Those excess electrons will bind to photoexcited electron-hole pairs, giving rise to negatively charged excitons (trions). Thus, the formation of trions will lead to the reduction of PL emissions of excitons A. On the other hand, applying a negative gate voltage will repel the excess electrons from surface towards the metallic contact (Fig. 4b). It will help to dissociate the trion into exciton A by promoting the extra electron in a trion to the conduction band edge. As a result, the exciton A emission is enhanced. Such different impacts of applying gate bias on the PL emissions both of exciton A and of trions can also be revealed by fitting the PL contribution of exciton A and trions and plotting the intensity ratios under different gate voltages from -30 V to 30 V, as shown in Fig. 4d which is extracted directly from the PL measurements under different gate bias (Fig. S5). It can be seen that applying negative bias increases the spectral weight of the exciton A while it is diminished under positive bias, which agrees very well with the previous interpretation. The PL modulation mechanism under the injection of carriers can be further confirmed by the excitation power dependent intensity ratios between trions and excitons A. It can be seen that the ratios between the PL intensity of the excitons A and trions decrease when increasing the excitation power, while both exciton A and trion emissions increase, as shown in Fig. 1(c, d) and summarized in Fig. S6. It indicates that higher pumping power will favor the formation of trions with respect to excitons A. This behavior is expected as more and more electrons are injected under higher excitation, binding to electron-hole pairs, and converting excitons into trions. Furthermore, since the cavity resonance of the gap mode metasurface is designed to match with the exciton A resonance, the modulation of exciton A resonance caused by the gate voltage will be remarkably amplified by the cavity effect of the GSP metasurface, leading to an ultra-sensitive tunability. This can be confirmed by investigating of the spectral dependent modulation depth in Fig. 4e. As it is shown, the spectral modulation curve shows remarkable wavelength selectivity with its peak at around 650 nm, corresponding to the exciton A-plasmon polariton resonance. Moreover, it is worthy to point out that other effects such as Pauli blocking, many-body interactions, and Stark effect have been widely found to cause shifts of the neutral exciton binding energy in the MOSFET structures under different voltage bias.[20,45] Thus, those effects are likely to modify the coupling between exciton A and GSP resonance in our structure as well, accounting for the modulation mechanism. In order to clarify it, the exciton A energy derived from the PL was investigated as a function of gate bias in Fig. 4f. Interestingly, it is found that the exciton A resonance in our structure almost keeps no change in wide range of gate bias from -30 V to 30 V. Considering that the GSP resonance of the metasurface has been designed to match to the exciton A resonance at 0 V, it implies that the existence of GSP metasurface helps to enhance the light-matter interactions inside the $MoS_2$ layer by making its excitonic resonance more robust to the electric static environment changes. The relatively small shift of the exciton resonance within our bias voltage range is beneficial to achieve significant tunability, which would otherwise comprehend the performance of the modulation mechanism investigated therein. Since the PL emissions of both exciton A and trion exhibit fast decay, typically within ps time scale, the modulation speed of our structure is mainly limited by the RC time constant of the circuit which is in the range of several nanoseconds (see supplementary for details). The operation speed of this structure can be further improved by shrinking the size of the metallic contacts.

**Experimental methods**

MoS$_2$ flakes were synthesized on SiO$_2$/Si substrates using MoO$_2$ (molybdenum dioxide) and sulfur powder in a quartz-tube-in-a-furnace vapor-phased deposition system at ambient pressure and then transferred onto the SiO$_2$/Au for metasurface integration (Fig. S7 and Fig. S8, see supporting information for details). The optical properties of the MoS$_2$ crystals were characterized by micro-photoluminescence (PL) spectroscopy, all the spectra were measured at room temperature (RT) using an inverted microscope coupled with an Andor spectrometer equipped with a intensified CCD camera. The excitation was provided by a temperature stabilized continuous (405 nm) laser, impinging onto the metasurface at normal incidence with a linear polarization oriented along the nanoridges.

**Conclusion**

Gate-tunable exciton-plasmon polariton light emitting metasurface is designed and demonstrated by integrating MoS$_2$ monolayer with gap plasmon mode-metasurfaces. This structure shows extraordinary advantages of controlling the light emitting characteristics such as polarization control, and selective enhancement of emission, benefitting from its unique resonant coupling between the excitonic resonance and the gap plasmon resonance. In particular, strong modulation of the polariton light emission can be realized by applying gate voltage. Interestingly, electrically tunable polaritonic light emitting metasurface could further leverage on the strong sensitivity of the MoS$_2$ d-orbitals participating in the indirect-to-direct transition, to both explore and advance the manipulation of light-matter interactions at the subwavelength dimensions, thus leading to novel nanoscale polaritonic optoelectronic devices.

## AUTHOR INFORMATION

**Corresponding Author:**

aluna@cinvestav.mx; Patrice.Genevet@crhea.cnrs.fr

**Author Contributions**

‡These authors contributed equally. The manuscript was written through contributions of all authors. All authors have given approval to the final version of the manuscript.

**Conflict of interest**

The authors declare that they have no conflict of interest.

**Acknowledgements and Funding:**

We acknowledge financial support from the European Union's Horizon 2020 under the European Research Council (ERC) grant agreement No. 639109 (project Flatlight). ADLB and ES acknowledge the financial support from SEP-CONACYT Ciencia Basica grant No 258674 and CONACYT-ERC grant 291826.

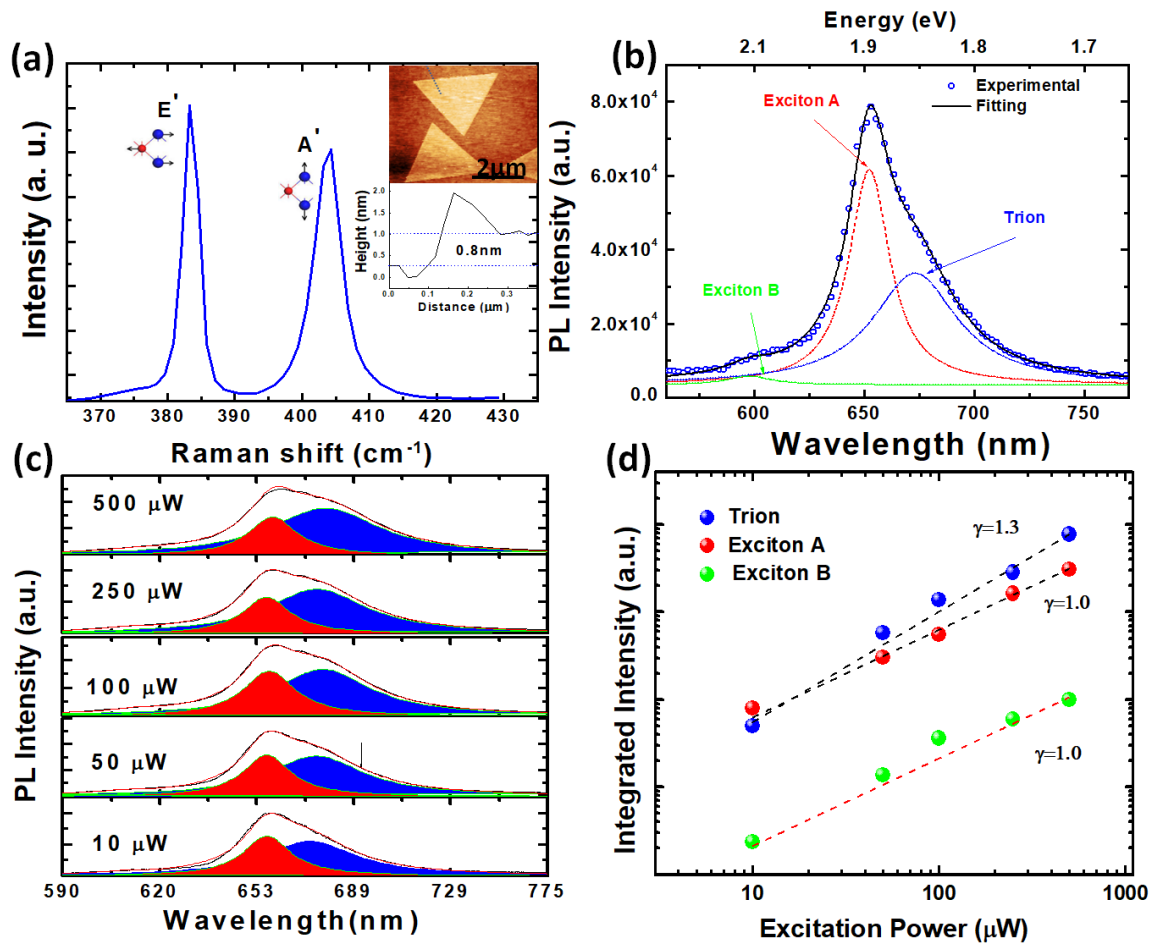

**Figure 1** (a) Raman spectrum of as-grown MoS$_2$ flakes on SiO$_2$ substrates. The energy difference between of E$^1_{2g}$ and A$_{2g}$ peaks and the thickness (~0.8 nm) obtain by AFM (inset) reveals the presence of MoS$_2$ monolayers (b) PL spectra of the MoS$_2$ transferred onto the Au/SiO$_2$ substrate. Note that the PL intensity and PL peaks of the MoS$_2$ on SiO$_2$/Au substrates have been tailored compared with as-grown MoS$_2$ on SiO$_2$/Si substrate and can be decomposed into three excitonic features. (c) and (c) the excitation-power dependent intensity of the major three PL components from the MoS$_2$ on Au/SiO$_2$ substrate, revealing the different nature of the excitonic transitions, respectively.

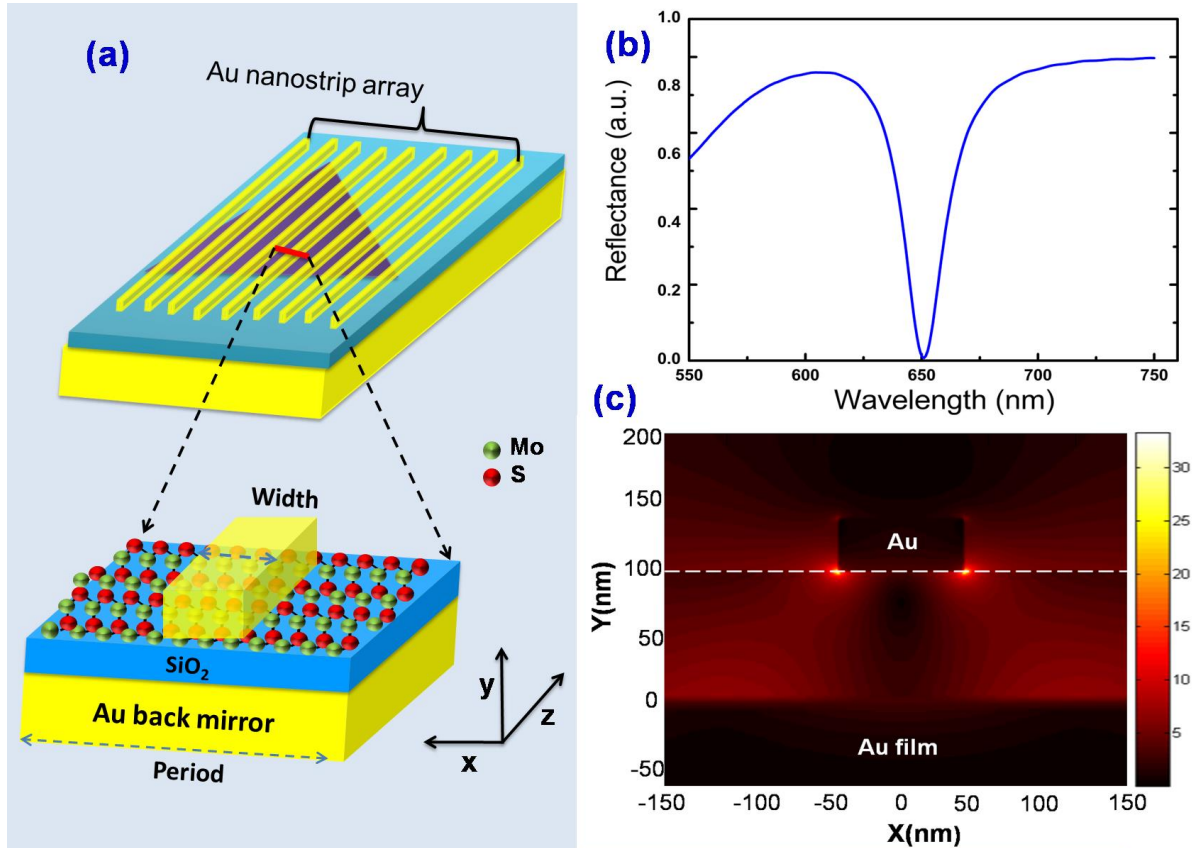

**Figure 2** (a) The schematic of the gap plasmon metasurfaces which comprises an array of Au metallic stripe and a Au substrate, separated by a 100 nm thick insulating $SiO_2$ dielectric layer and a monolayer $MoS_2$ flakes (b) The simulated reflectance of the bare Gap Plasmon metasurface when the width of the Au nanostrip is fixed at 72 nm and the spacing between the nanostrip is fixed at 373 nm. (c) The calculated electric field intensity in the plane transverse to the metallic wires showing that the electric field is tightly confined within the gap layer, and the dash line indicates the position of $MoS_2$.

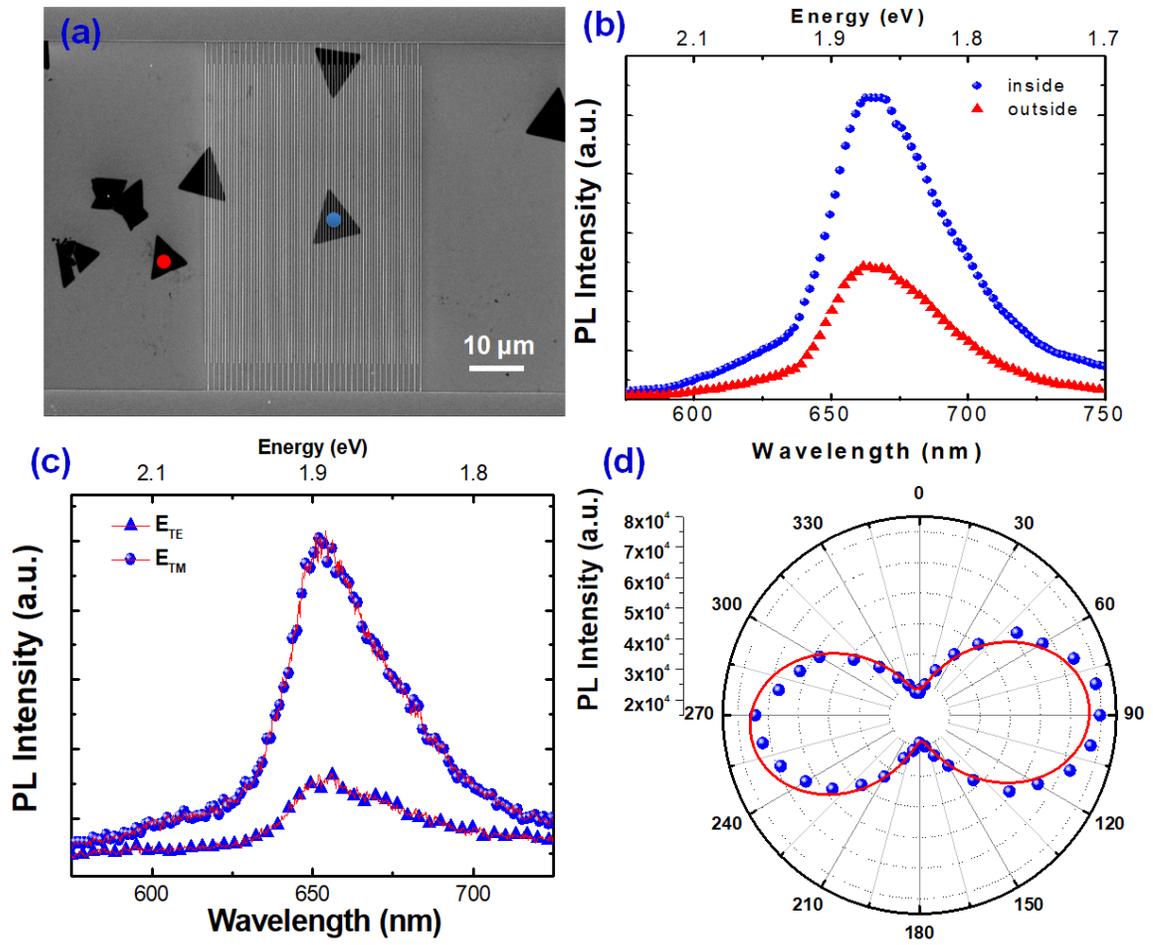

**Figure 3** (a) The SEM image of the fabricated hybrid MoS2 gap-mode metasurface; (b) shows the PL emissions of MoS2 both inside and outside the metasurface at location denoted in by the blue and red dots respectively; (c) The polarization dependent light emission of MoS2 inside the metasurface; (d) shows the polar plot of emission.

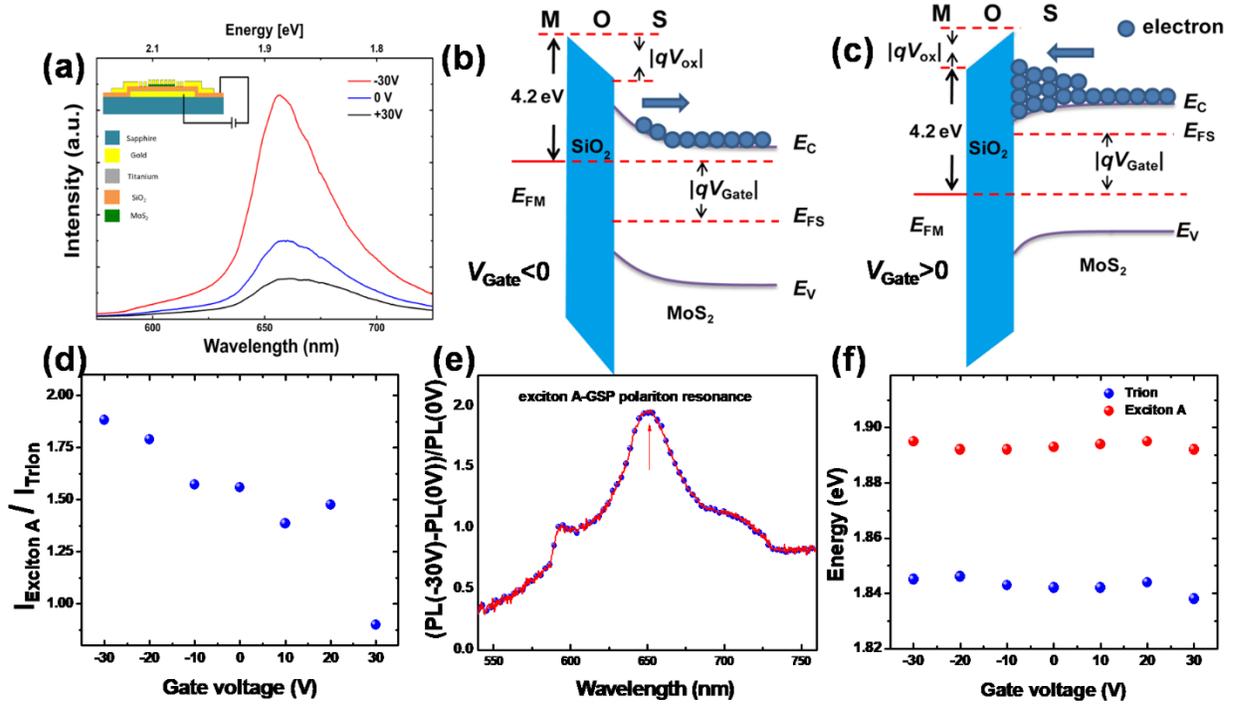

**Figure 4** (a) PL spectra of gated MoS$_2$-gap plasmon metasurfaces under different applied voltages, indicating strong gate tunable emissions, the inset shows the schematic representation of the fabricated devices. The band diagram of the MoS$_2$/SiO$_2$/Au structure, at negative (b) and positive gate bias (c) are used to explain the tuning process between exciton A and trion. Positive gate bias will cause excess electrons accumulated at the interface by raising the Fermi level while the negative bias can expel the excess electrons from the surface into the metallic contact (d) the ratio of emission intensity between the exciton A and trion, showing change of the spectral weight under different bias (e) the relative change of light emission at -30V with respect to that at 0 V, indicating the selectively enhancement of modulation depth (f) The energies of exciton A and trion as a function of gate bias.

**Supplementary materials**

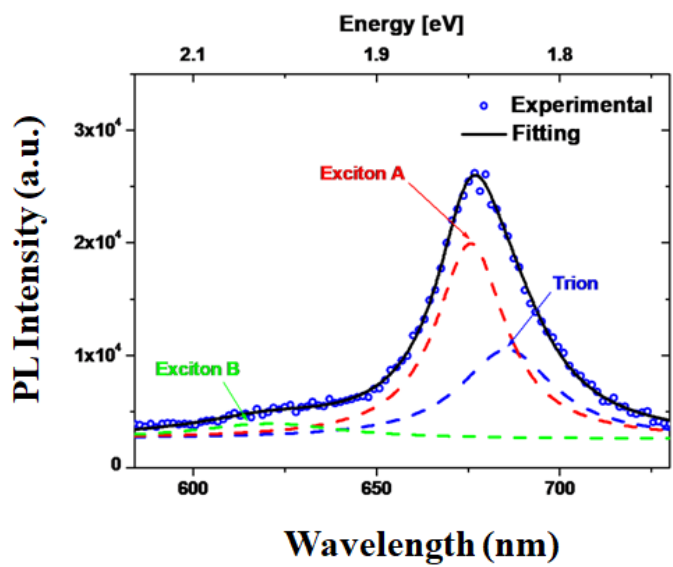

**Figure S1.** PL spectra of the as-grown $MoS_2$ on the $SiO_2$/Si substrate, which can be decomposed into three excitonic features, corresponding to exciton A, exciton B and trion emission, respectively.

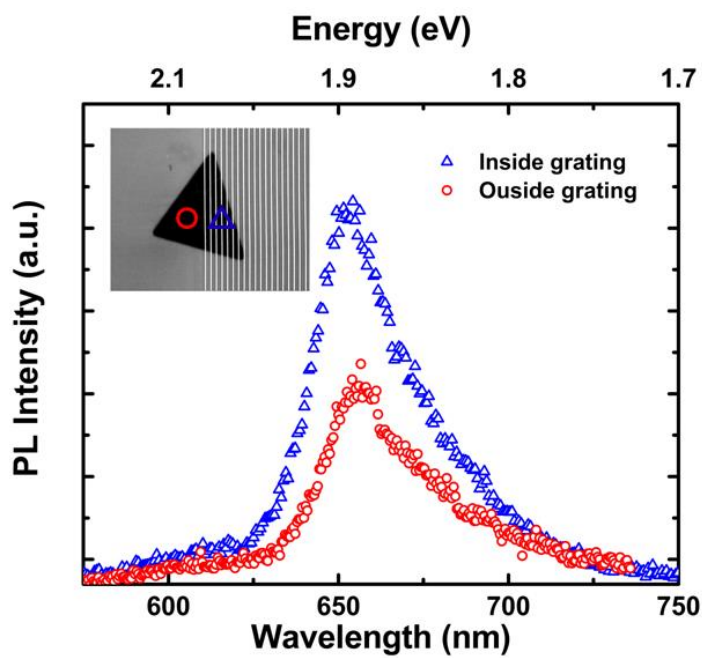

**Figure S2.** PL spectra of the $MoS_2$ on the $SiO_2$/Au substrate inside and outside the grating structure, confirming the large luminescence enhancement due to the surrounding of metallic grating.

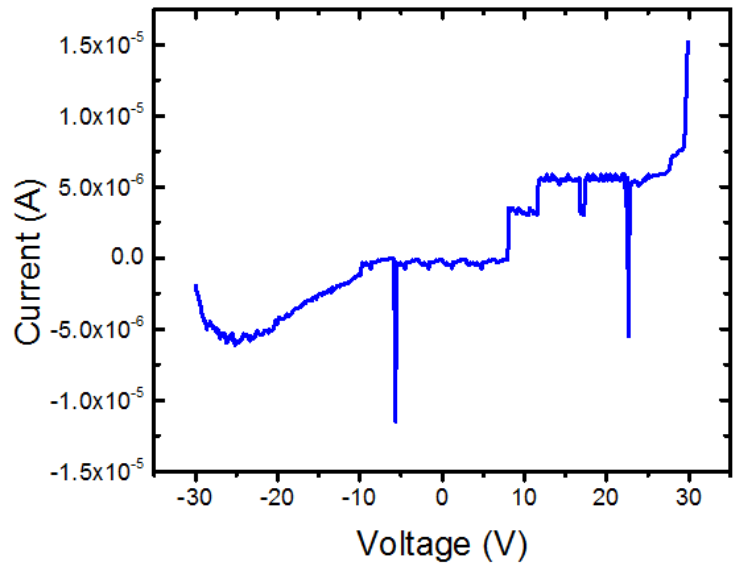

**Figure S3**. I-V characteristic of the structure as a function of gate voltages, demonstrating the good dielectric property of the SiO$_2$ layer with negligible leakage current.

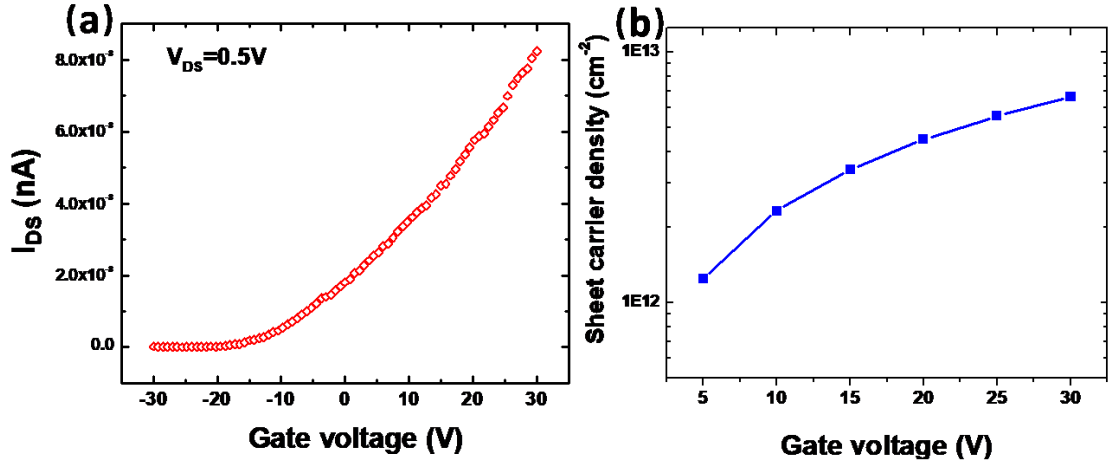

**Figure S4.** We estimate the carrier density change by directly measuring the electrical properties of the as-grown MoS$_2$ flakes. To this end, we fabricated a MoS$_2$ based FET device on 100nm SiO$_2$ on highly doped *p*-type, using Si as a back gate (drain and source electrodes (Ti/Au) were deposited by optical lithography directly on MoS$_2$ single layer). (a) depicts the transfer characteristics of the MoS$_2$ FET. The I$_{ds}$-V$_{gs}$ curve recorded at V$_{DS}$=0.5V confirms the *n*-type nature of the MoS$_2$ flakes, by applying positive V$_{gate}$ bias the transistor enters in electron accumulation regime whereas for negative V$_{gate}$ bias the channel is depleted. Using the parallel-plate capacitor model for the carrier density concentration $n \approx \frac{C_G}{q}(V_G - V_T)$ and taking the threshold voltage as V$_T$ = -0.8V and C$_G$=ε$_0$ε$_r$/d$_{ox}$, where ε$_0$ is the vacuum dielectric constant and ε$_r$ the SiO$_2$ relative dielectric constant of 3.9; carrier density as a function of gate bias is determined (b).

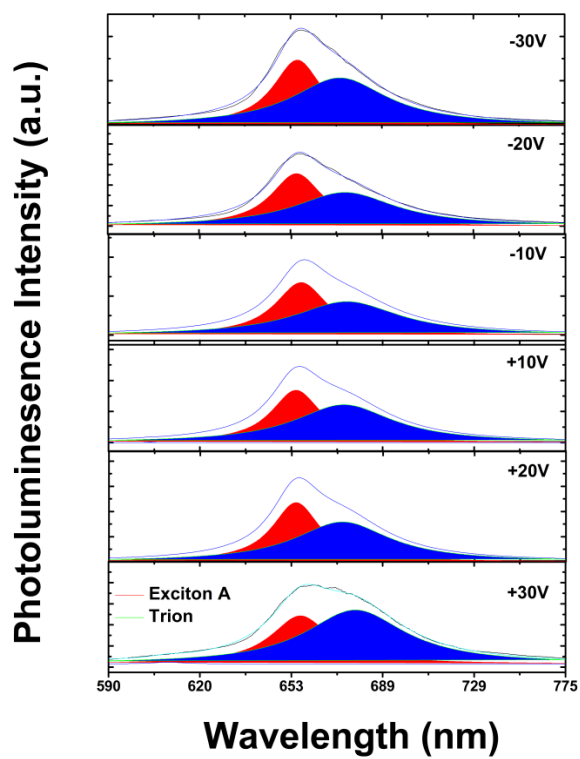

**Figure S5.** PL emissions from MoS$_2$ transferred onto SiO$_2$/Au substrate as a function of gate voltages.

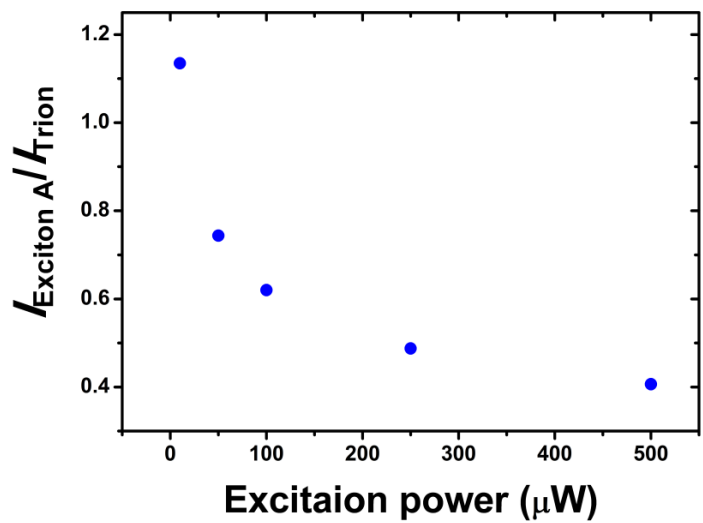

**Figure S6.** Dependence of trion/ exciton PL intensity ratios on the excitation power.

**Modulation speed:** The rise time of the structure estimated by its RC time constant ($\tau_r=2.2\tau=2.2RC$) is in the range of several nanoseconds based on the load resistance $R$ and the capacitance of the circuit $C=\varepsilon_0\varepsilon_r A/d_{ox}$, where $\varepsilon_0$ is the vacuum dielectric constant, $\varepsilon_r$ is the SiO$_2$ relative dielectric constant of 3.9, $A$ is the size of the capacitor surface and d is the thickness of SiO$_2$ dielectric layer. Therefore, it can be seen that the operation speed of the structure is mainly limited by its RC constant. Note that the RC time constant of the current structure in this work is mainly limited by the large metallic contacts which are used for wire bonding with respect to the subwavelength size of grating. Thus, the operation speed of this structure can be further improved by shrinking the size of the metallic contacts.

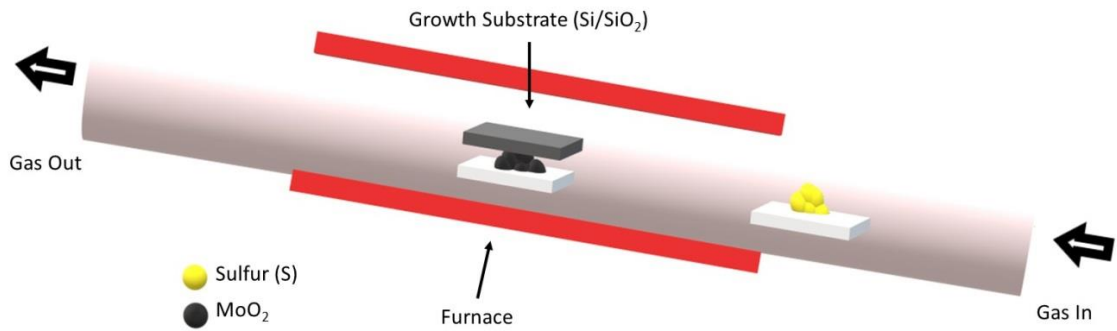

**Figure S7.** Schematic of CVD growth of MoS$_2$

**Growth details:** Single MoS$_2$ flakes were synthesized by atmospheric chemical vapor deposition APCVD method at 700°C in a 1in quartz tube with argon as a carrier gas. MoO$_3$ (99% Sigma Aldrich) and Sulfur powders (99% Sigma Aldrich) are used as precursors, while MoO$_3$ is placed at the center of the furnace in an alumina boat, sulfur is situated upstream at the edge of the furnace. Finally, SiO$_2$/Si substrates are cleaned using isopropanol and acetone and placed face down in the alumina boat containing the MoO$_3$ powder. The temperature of the system is first increased from room temperature to 300°C with a rate of 10°C/min, after 30 minutes the temperature is raised up to 700°C with a ramp of 30°C/min, the system is immediately cooled down naturally once it reaches the growth temperature (700°C).

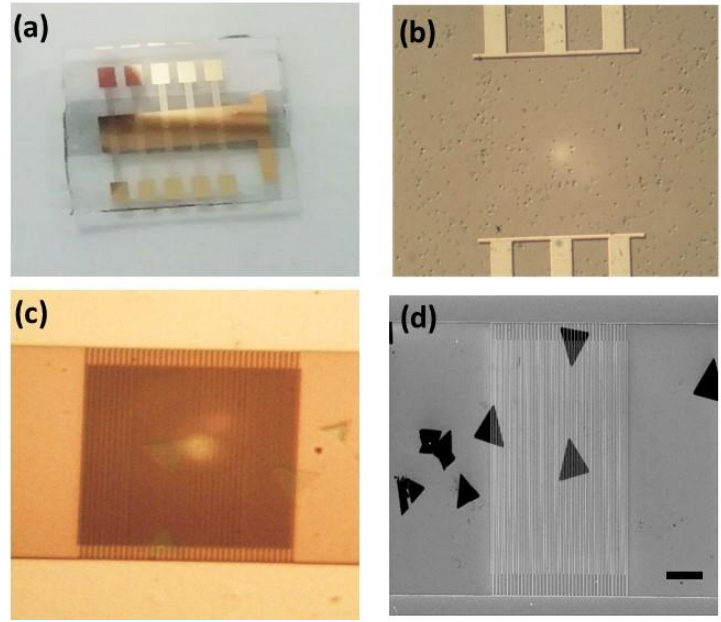

**Figure S8**. Fabrication process of the gap plasmon metasurface. a) Photograph of the device, b) MoS$_2$ flakes transferred on SiO$_2$/Au substrates. c) and d) optical and SEM images after e-beam lithography and lift off processes

**Fabrication details:** First, to form the metallic mirror of our devices, 1cm x 1cm sapphire substrates were covered using thermal tape leaving a window of 3mm x 10mm to deposit a layer of Ti/Au (5/200nm thick). Then samples are introduced in a conventional sputtering system to deposit 100nm of SiO$_2$ and the tape is removed. An optical lithography step is then performed to define large Ti/Au (10/ 200nm) metallic contacts. MoS$_2$ flakes are transferred from SiO$_2$ substrates using PMMA and KOH solution on the Au/SiO$_2$ sandwich (a, b). Finally, metallic nanostrip-array metasurface was defined by means of e-beam lithography followed by a 50 nm Au film deposited by e-beam evaporation and lift-off during night on PG remover at 60°C (c, d).